\documentclass[9pt,conference]{IEEEtran}

\usepackage[preprint]{waspaa25}

\usepackage{bm} 
\usepackage{algorithm}
\usepackage{algpseudocode}
\usepackage{amsmath}
\usepackage{siunitx}
\usepackage{multirow}
\usepackage{caption}
\usepackage{stfloats}

\title{Hybrid-Sep: Language-queried audio source separation via pre-trained Model Fusion and Adversarial Consistent Training}

\name{Jianyuan Feng$^{1}$,
      Guangzheng Li$^{1}$,
      Yangfei Xu$^{1}$}
\address{$^{1}$ByteDance, China
}

\begin{document}

\maketitle

\begin{abstract}
Language-queried Audio Separation (LASS) employs linguistic queries to isolate target sounds based on semantic descriptions. However, existing methods face challenges in aligning complex auditory features with linguistic context while preserving separation precision. Current research efforts focus primarily on text description augmentation and architectural innovations, yet the potential of integrating pre-trained self-supervised learning (SSL) audio models and Contrastive Language-Audio Pretraining (CLAP) frameworks, capable of extracting cross-modal audio-text relationships, remains underexplored. To address this, we present HybridSep, a two-stage LASS framework that synergizes SSL-based acoustic representations with CLAP-derived semantic embeddings. Our framework introduces Adversarial Consistent Training (ACT), a novel optimization strategy that treats diffusion as an auxiliary regularization loss while integrating adversarial training to enhance separation fidelity. Experiments demonstrate that HybridSep achieves significant performance improvements over state-of-the-art baselines (e.g., AudioSep, FlowSep) across multiple metrics, establishing new benchmarks for LASS tasks. Demo can be find in web.\footnote{Demo available at: \url{https://windval.github.io/Hybrid-Sep-Demo}}
\end{abstract}

\section{Introduction}
\label{sec:intro}

Traditional audio source separation methods typically require specialized models for each predefined sound category, fundamentally limiting the scalability of separable sources. This approach faces practical challenges, as real-world audio scenes often contain indeterminate numbers of overlapping sources. For instance, street noise may simultaneously include vehicle sounds, speech fragments, and footstep patterns. Language-queried audio source separation (LASS) has emerged as a promising solution that enables flexible target specification through natural language descriptions. Pioneering work like AudioSep \cite{liu2024separate} uses a text encoder from contrastive language-audio pretraining (CLAP) to convert language queries to audio-aware embeddings, subsequently employing a ResUNet architecture for source separation conditioned on these embeddings. Subsequent studies have enhanced caption diversity through prompt engineering \cite{yin2025exploring}  \cite{lee2024performance}. while Yin et al. \cite{yin2025exploring} improved robustness by retrieving text embeddings from a large language model (LLM)-generated cache and introducing controlled noise injection into the text embedding in training.   Building upon the principle that separated components should reconstruct the original mixture, Chung et al.\cite{chung2024language} incorporated contrastive learning into LASS frameworks. Recent advancements like FlowSep \cite{yuan2025flowsep} introduce an adversarial diffusion framework to LASS and show significant improvement in semantic-similarity-based matrices such as CLAP-score \cite{xiao2024reference} and FAD \cite{kilgour2018fr}, this approaches face challenges in preserving signal amplitude fidelity and require computationally intensive multi-step iterative sampling processes. To address these shortcomings, the recent development in adversarial diffusion training (ADT) may be worth exploring. Consistency model \cite{song2023consistencymodels} which provides a training method that trains the diffusion model object to enforce temporal consistency along the probability flow ODE (PF-ODE) trajectory:

\begin{equation}
  \label{eq:wave_equation}
\min_{\theta} \mathbb{E}_{x_{t_{n}}, x_{t_{n+1}}} \left[ \mathcal{D}\left( f_\theta(x_{t_{n+1}}, t_{n+1}), f_\theta(x_{t_n}, t_n) \right) \right]
\end{equation}

where $\mathcal{D}$ is a distance metric, $x_{t_{n+1}}$ , ${x_{t_n}}$ , are adjacent points sampled from the same PF-ODE trajectory, $f_\theta$ maps any noisy sample $x_{t}$ to a consistent estimate $x_{\epsilon}$. With this consistency constraint formulation, diffusion models can achieve similar results with much fewer diffusion steps. Tianhong et al. \cite{li2024autoregressiveimagegenerationvector} proved that introducing a small conditional denoise model acts as a distributional guide to enforce the autoregressive model’s predictions to align with the continuous-valued data manifold. This approach decouples the training-time noise prediction task from inference-time generation, enabling fast single-step sampling while retaining diffusion-like quality.
 
The CLAP framework establishes audio-text alignment through contrastive learning between joint embedding spaces. CLAP shows an advantage in sematic-level audio tasks such as audio auto caption (AAC) \cite{li2025drcap}, and sound event detection (SED)\cite{wu2023large}. Text-conditioned approaches encounter inherent ambiguity in that single text descriptions may correspond to numerous acoustic realizations, creating optimization challenges for separation networks. In contrast, CLAP's audio encoder produces deterministic embeddings for specific audio samples. The quantitative evaluation of the LASS models \cite{yuan2025flowsep} demonstrates a pronounced cross-modal discrepancy, with separated audio embeddings exhibiting high cosine similarity to target audio embeddings (exceeding 80\%) while maintaining low alignment scores (0.4\%) compared to query text embeddings in the CLAP latent space.  Recent temporal-aware extensions like T-CLAP \cite{yuan2024t} further improved time domain semantic feature extraction by augmenting time-awarded audio capture data and temporal-focused contrast loss. As a result, using target audio embedding or predicted target audio embedding as the condition of the separation model may be worth exploring.

Self-supervised learning (SSL) methods in audio analysis aim to derive robust acoustic representations through self-supervised tasks. HuBERT \cite{hsu2021hubert} adapts masked language modeling to speech signals by predicting discrete pseudo-labels (k-means on MFCC/acoustic features) from masked acoustic features, while BEATs \cite{chen2022beats} merged the acoustic tokenizer into SSL and perform spectrogram masked token prediction as the SSL tasks which achieve SOTA in SED tasks. However, the SSL effect in audio source separation is not fully explored. MERT \cite{li2023mert}, applies HuBERT SSL training framework and takes Constant-Q transform (CQT) rather than MFCC as the audio feature extractor. By adding a simple downstream model instead of fine-tuning the pre-trained model MERT shows improvement in music understanding tasks such as music tagging and key detection, but the reported result is much worse compared with SOTA models in audio separation tasks. 

In this work, we introduce Hybrid-Sep, a 2-stage LASS model that takes advantage of both CLAP and SSL pre-trained models. A novel adversarial consistent training (ACT) method is introduced to improve the performance of LASS tasks without suffering from the heavy computational costs of multi-step inference. Ablation experiments were conducted to illustrate the effects of pre-trained models and the ACT. Hybrid-Sep is compared with other current state-of-the-art (SOTA) LASS models under both free-form description and categorical tag query conditions. Experimental results demonstrate consistent improvements among all metrics: Signal-to-Distortion Ratio (SDR), CLAPscore, and Fréchet Audio Distance (FAD).

\section{PROPOSED METHOD}
\label{sec:PROPOSED_MODEL}
\begin{figure}[t]
  \centering
  \centerline{\includegraphics[width=\columnwidth]{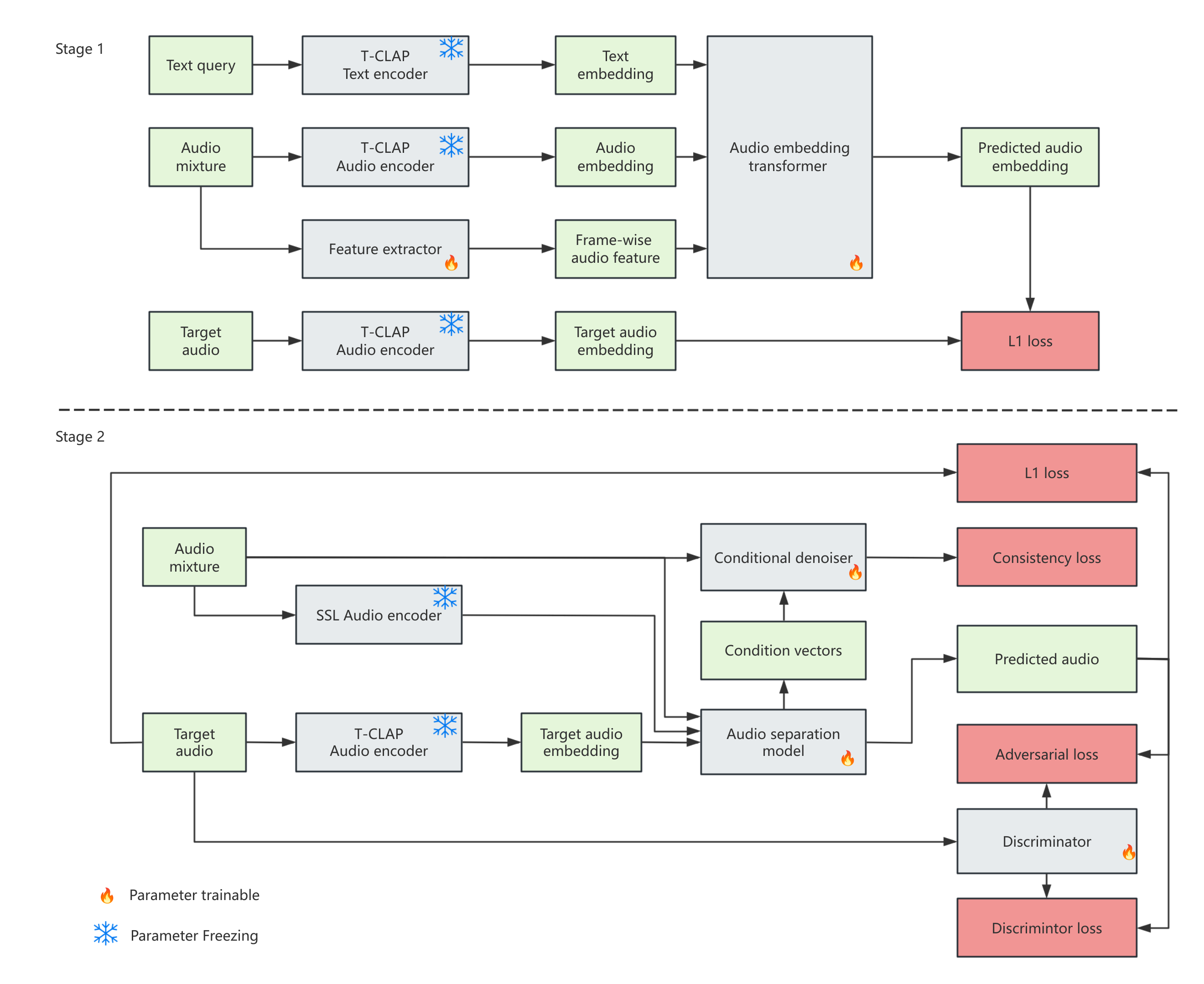}}
  \caption{2 stage training of the Hybrid-Sep}
  \label{fig:train}
\end{figure}

\begin{figure}[t]
  \centering
  \centerline{\includegraphics[width=\columnwidth]{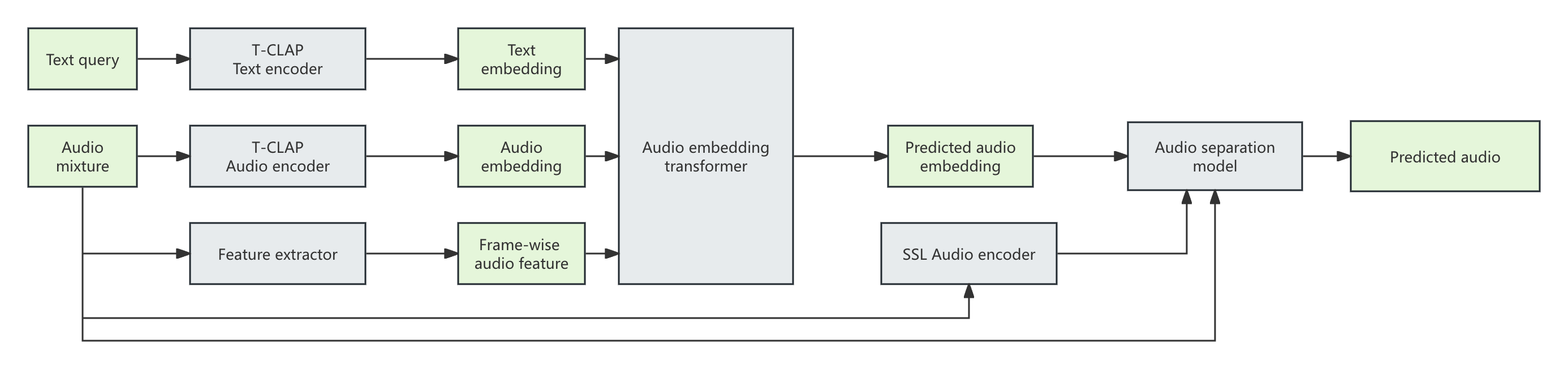}}
  \caption{Inference pipeline of the Hybrid-Sep}
  \label{fig:inference}
\end{figure}

By decomposing the LASS tasks into target embedding prediction followed by embedding-conditioned separation, the model establishes a latent bottleneck representation that forces disentanglement of linguistic semantics from temporal characteristics, effectively mitigating error propagation caused by imperfect text-audio correspondence. The overlook of the training and inference process of the Hybrid-Sep is illustrated in \Cref{fig:train} and \Cref{fig:inference} respectively. 

\subsection{Audio embedding transformer}

In the first stage, the audio embedding is trained to predict the required audio embedding from the corresponding text query and audio mixture. The pre-trained T-CLAP text encoder converts the text query to text embedding. The audio embedding represents the semantic features and the frame-wise acoustic features are achieved through a T-CLAP audio encoder and a feature extractor (FE) respectively. Then an audio embedding transformer (AET) is designed to project the text embedding to target audio embedding. The FE contains an STFT module followed by 3 transformer layers, and the audio embedding transformer is simply 32 transformer layers. The target audio embedding is constrained to the corresponding audio embedding extracted from the target audio using a T-CLAP audio encoder and L1 loss. The parameters in the audio encoder and text encoder of T-CLAP are frozen, only parameters in the FE and audio embedding transformer require training.

\subsection{Adversarial consistent training}

In the second stage, the audio separation model (ASM) is designed to separate the target audio from the mixture audio based on target audio embedding. The training process follows a novel adversarial consistent training (ACT) schedule which includes 3 main parts. Firstly, a L1 loss is used to constrain the distance between the predicted and target audio. Secondly, a GAN-based method using a multiscale spectrogram discriminator to reduce the artifacts, the structure of the discriminator and loss design remains the same as it is in Xcodec\cite{défossez2022highfidelityneuralaudio}. Finally, a conditional denoising module is introduced to enforce noise invariance, ensuring that the separated audio predictions remain consistent across different noise-corrupted mixtures. The training steps of ACT are summarized as follows:

\begin{algorithm}
\caption{Training steps of ACT}
\label{alg:gan_audio_consistency}
\begin{algorithmic}
\Require Mixed audio batch $\mathbf{X}_{\text{mixed}}$, target audio batch $\mathbf{X}_{\text{target}}$, Audio separation model $ASM$, Discriminator $D$, Conditional Denoiser $CD$, total iterations $N$, L1 loss weight $\lambda_{\text{L1}}$, step schedule $N(\cdot)$, EMA decay rate schedule $\mathbf{\sigma}(\cdot)$, $d(\cdot, \cdot)$, $\lambda_{\text{consist}}(\cdot)$

\State Initialize weights of $ASM$, $D$, and $CD$
\State Define optimizers: $\text{opt}_{ASM}$ (for $ASM$ and $CD$), $\text{opt}_D$ (for $D$)

\For{iteration $k = 1$ \textbf{to} $N$}
    \State \textbf{Phase 1: Train $D$}
    \State Generate separated audio: $\mathbf{X}_{\text{gen}} \gets ASM(\mathbf{X}_{\text{mixed}})$
    \State Compute $\mathbf{d}_{\text{real}} \gets D(\mathbf{X}_{\text{target}})$, $\mathbf{d}_{\text{fake}} \gets D(\mathbf{X}_{\text{gen}})$
    \State $\mathcal{L}_D \gets \frac{1}{2}\left(\mathbb{E}[(\mathbf{d}_{\text{real}} - 1)^2] + \mathbb{E}[\mathbf{d}_{\text{fake}}^2]\right)$
    \State Update $D$: $\text{opt}_D.\text{step}(\nabla_D \mathcal{L}_D)$
    
    \State \textbf{Phase 2: Train $ASM$ and $CD$}
    \State Regenerate audio: $(\mathbf{X}_{\text{gen}}, \mathbf{c}) \gets ASM(\mathbf{X}_{\text{mixed}})$

    \State $\mathbf{d}_{\text{fake}} \gets D(\mathbf{X}_{\text{gen}})$
    \State Sample noise: $\mathbf{\epsilon_1}$, $\mathbf{\epsilon_2} \sim \mathcal{N}(0, \sigma(k))$
    \State Get noisy targets: $\mathbf{X}_{\text{noisy1}} \gets \mathbf{X}_{\text{target}} + \mathbf{\epsilon_1}$, $\mathbf{X}_{\text{noisy2}} \gets \mathbf{X}_{\text{target}} + \mathbf{\epsilon_2}$
    \State Denoise: $\mathbf{X}_{\text{pred1}} \gets CD(\mathbf{X}_{\text{noisy1}}, \mathbf{c})$, $\mathbf{X}_{\text{pred2}} \gets CD(\mathbf{X}_{\text{noisy2}}, \mathbf{c})$
    \State Consistency loss: $\mathcal{L}_{\text{consist}} \gets d[\mathbf{X}_{\text{pred1}} , \mathbf{X}_{\text{pred2}}.\text{detach()}]$
    
    \State L1 loss: $\mathcal{L}_{\text{L1}} \gets \|\mathbf{X}_{\text{gen}} - \mathbf{X}_{\text{target}}\|_1$
    \State Adversarial loss: $\mathcal{L}_{\text{adv}} \gets \mathbb{E}[(\mathbf{d}_{\text{fake}} - 1)^2]$
    \State Total loss: $\mathcal{L}_T \gets \mathcal{L}_{\text{adv}} + \lambda_{\text{L1}} \mathcal{L}_{\text{L1}} + \lambda_{\text{consist}}(k) \mathcal{L}_{\text{consist}}$
    \State Update $ASM$ and $CD$: $\text{opt}_{ASM}.\text{step}(\nabla_{ASM, CD} \mathcal{L}_T)$
\EndFor
\State \Return $ASM^*$, $D^*$, $CD^*$
\end{algorithmic}
\end{algorithm}

\begin{figure}[t]
  \centering
  \centerline{\includegraphics[width=\columnwidth]{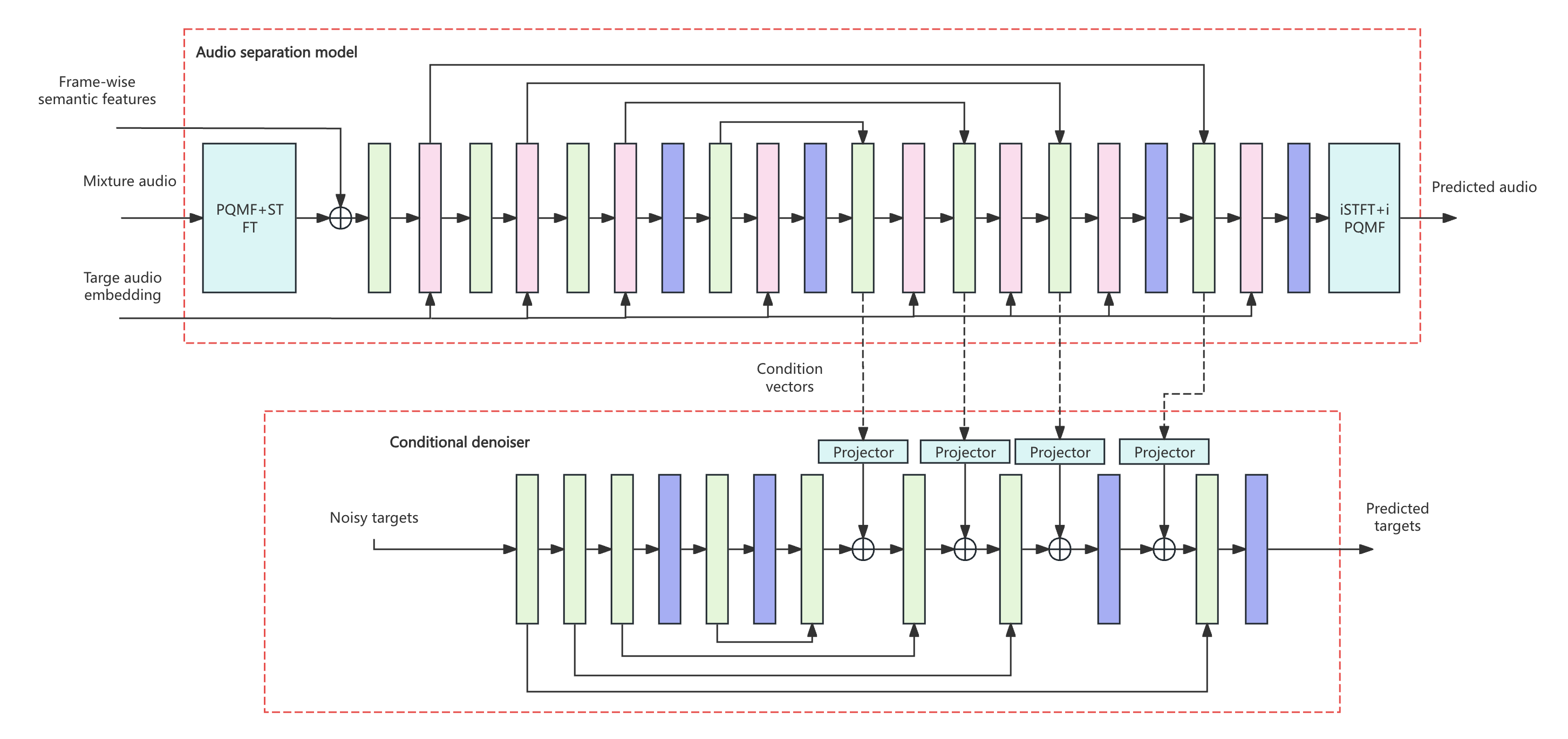}}
  \caption{Model structure of audio separation model and conditional denoiser (Green, red, and blue blocks are indicating the TRCNN, FA, and TEACA layer respectively)}
  \label{fig:detailed_audio_separation}
\end{figure}

For Hybrid-Sep training the $\lambda_{\text{L1}}$ is set to 1. The exponential moving average (EMA) schedules ($\lambda_{\text{consist}}(k)$ and $\sigma(k)$) are used to generate consistency loss which helps to improve the final performance and stabilize the training process which set the same as the consistency model experience \cite{song2023consistencymodels}\cite{song2023improvedtechniquestrainingconsistency}.\footnote{\url{https://github.com/openai/consistency_models}} The detailed model structures of the ASM and the CD are illustrated in \Cref{fig:detailed_audio_separation}.

ASM first uses PQMF \cite{galand1983design} to evenly split the audio mixture into 4 subbands, and convert to spectrograms using a short-time Fourier transform (STFT). These spectral inputs are concatenated with frame-wise semantic features extracted by a 90M-parameter BEATs encoder \cite{chen2022beats} pre-trained via SSL, forming a hybrid representation that jointly encodes acoustic and high-level contextual information. To obtain the predicted sub-band spectrograms, the composite feature is subsequently processed through an interleaved stack of three types of modules: 

1. Temporal recurrent convolutional neural network (TRCNN) modules for multi-scale temporal pattern extraction which stack a CNN layer (Using Conv2d in the first four TRCNNs and Deconv2d in the last four TRCNNs), a GELU layer, and a BiLSTM layer. 

2. Frequency-axis attention (FA) modules which contain a conformer layer performing conformer-based spectral recalibration along the frequency dimension.

3. Target-embedding-aware cross-attention (TEACA) modules that apply channel-adaptive attention using target audio embeddings as query vectors. 

Then the predicted sub-band spectrograms are processed with inverse STFT and iPQMF synthesis to reconstruct the target waveform. Meanwhile, feature maps from the final four TRCNN modules are projected via dedicated linear transformers and channel-concatenated to form condition vectors for the CD, which is a streamlined variant of the main separation architecture that excludes TAECA modules while retaining the same structured TRCNN and FA components for noise-robust refinement. This hierarchical design enables coordinated feature learning while maintaining parameter efficiency. The detailed hyper-parameters settings of the Hybrid-Sep are listed in \cref{tab:module_config}.

\section{Experiment and result}
\subsection{Training setting}
The query text of the LASS task can be a detailed caption such as "the fireworks are soaring into the sky and exploding" or a simple keyword like "fireworks". We take both scenarios and train the Hybrid-Sep with audio-keyword paired data and audio-caption paired data. 

The audio-keyword paired dataset for training including the training data of AudioSet \cite{gemmeke2017audio}, training data of MUSDB18HQ \cite{rafii2017musdb18}, training data of Slakh \cite{manilow2019cutting}, noise and speech data from DNS Challange \cite{dubey2023icassp}. The audio-caption paired dataset for training including Audiocaps \cite{audiocaps},WavCaps \cite{mei2024wavcaps},Clotho V2 \cite{drossos2019clothoaudiocaptioningdataset}, FSD50K \cite{fonseca2021fsd50k}, and Auto-ACD \cite{sun2024auto}. Overall around 150K hours of audio samples and 2.1M audio queries were used for training.

The audio in all datasets has been sampled to 10s chunk size audio and resampled to 44.1kHz. Keywords such as "Guitar" will be directly used as a language query. When multiple keywords were present (as in AudioSet), they were combined into a single phrase (e.g., "guitar, bird and speech"). If there are multiple captions for one audio sample, one caption will be randomly selected as the query. 

We implement two distinct audio mixture augmentation protocols for audio-keyword paired and audio-caption datasets. For audio-keyword paired datasets, except for AudioSet, individual samples mainly contain single audio sources, where mixtures are synthesized by combining 2–4 samples. The target audio is constrained to exhibit distinct keywords from all non-target segments within each mixture to enforce discriminative separation. For audio-caption paired datasets, mixtures are generated by blending two samples. To avoid source contamination, audio-keyword-paired data is prohibited from being mixed with caption-paired data during augmentation. The energy ratio between target and non-target components is uniformly sampled within the range from $-20 dB$ to +$20 dB$, ensuring a balanced representation of relative signal dominance across training instances.

\begin{table}[ht]
\centering
\caption{Hyper-parameters setting for Hybrid-Sep}
\label{tab:module_config}
\begin{tabular}{lll}
\toprule
\textbf{Module} & \textbf{Layers} & \textbf{Parameters} \\ 
\midrule
\multirow{1}{*}{FE} 
 & Transformer*3 & \begin{tabular}[c]{@{}l@{}}
                  Heads: 8, Dim: 512
                \end{tabular} \\ 
\midrule

\multirow{1}{*}{AET} 
 & Transformer*32 & \begin{tabular}[c]{@{}l@{}}
                    Heads: 8, Dim: 512 \\
                  \end{tabular} \\ 
\midrule

\multirow{3}{*}{ASM/CD} 
 & TRCNN & \begin{tabular}[c]{@{}l@{}}
              CNN-Kernels: (1,1)×8 \\
              CNN-Strides: [2,2,2,2,2,2,2,2] \\
              CNN-Channels: [48,96,192,384,192,96,48,8]\\
              BiLSTM-dim: [12,24,48,96,48,24,2]
            \end{tabular} \\ 
\cmidrule(lr){2-3}
 & FA*4 & \begin{tabular}[c]{@{}l@{}}
            Kernal size: 9\\
        \end{tabular} \\ 
\cmidrule(lr){2-3}
 & TEACA & \begin{tabular}[c]{@{}l@{}}
            Dim: 64
            Attention Heads: 8 
          \end{tabular} \\ 
\bottomrule
\end{tabular}
\end{table}
Hybrid-Sep comprises approximately 460M trainable parameters: 40M in FE, 415M in AET, and 25M in ASM. For all the experiments, 64 V100 GPUs with 32G as memory are used to train the Hybrid-Sep. The optimization employs AdamW\cite{loshchilov2017decoupled} with an initial learning rate of $10^{-4}$,$\beta_1$=0.8  and $\beta_2$= 0.99. Both training stage 1 and stage 2 take about 60000 steps to converge with 128 as the batch size. The single batch inference real-time factor (RTF) is 0.003 as measured on a single V100 GPU. 

\subsection{Experiment setting}
Both caption-audio paired and keyword-audio paired testing data are generated to evaluate the performance of the Hybried-Sep. The caption-audio paired testing including DCASE synthetic validation data (DE-S) includes 3000 synthetic mixtures with signal-to-noise-ratio (SNR) between -15dB to 15dB, and DCASE real-world evaluation data (DE-R) containing 200 audio samples collected from real-world recording. The keyword-audio paired testing data is generated by randomly mixing the evaluation dataset of Audioset, MUSDB18, and Slakh, in total 600 mixtures were generated and each mixture contains 4 audio samples (only 1 target audio).

Hybrid-Sep is compared with two available SOTA LASS models including AudioSep and FLowSep with inference steps as 10. To illustrate the effects of 2-stage training, the SSL audio encoder integration, and ACT, 3 versions of Hybrid-Sep model are developed. The models including Hybrid-Sep 1 stage version using text embedding (extract from text query using T-CLAP text encoder), Hybrid-Sep without SSL audio encoder, and Hybrid-Sep without CD are trained and tested.

From our knowledge, Hybrid-Sep is the first LASS model capable of working in a full-band environment. The maximum working sampling rate of AudioSep and FlowSep are 32kHz and 16kHz respectively. To ensure methodological rigor, the comparative experiments across different models were consistently conducted at a uniform sampling rate of 16 kHz, while the ablation studies were performed at a higher fidelity sampling rate of 44.1 kHz.

\subsection{Result}

\begin{table*}[t]
\centering
\caption{Comparing performance of LASS models on testing datasets. (KA is short for keyword-audio paired testing data)}
\label{tab:table1}
\sisetup{
    table-format=2.2,  
    table-number-alignment=center,
    tight-spacing=true,
    reset-text-series = false, 
    text-series-to-math = true
}
\begin{tabular}{l
    *{2}{S[table-format=2.2]}
    *{2}{S[table-format=2.2]}
    *{3}{S[table-format=2.1]}  
    *{2}{S[table-format=2.1]}  
    *{2}{S[table-format=1.3]}  
    }
    \toprule
     \multicolumn{1}{c}{} & 
    \multicolumn{2}{c}{\textbf{SDR} $\uparrow$} & 
    \multicolumn{2}{c}{\textbf{SDRi} $\uparrow$} & 
    \multicolumn{3}{c}{\textbf{CLAPscore} $\uparrow$} & 
    \multicolumn{2}{c}{\textbf{CLAPscoreA} $\uparrow$} & 
    \multicolumn{2}{c}{\textbf{FAD} $\downarrow$} \\
    \cmidrule(lr){2-3} \cmidrule(lr){4-5} \cmidrule(lr){6-8} \cmidrule(lr){9-10} \cmidrule(lr){11-12} 
    &{DE-S} &{KA}&{DE-S} &{KA} &{DE-S} &{DE-R} &{KA} &{DE-S} &{KA} &{DE-S} &{KA}\\
    \midrule
    No Processing (16kHz) &  0.03 & 0.81&{-} & {-}& 23.2 & 22.7 & 22.3& 71.3  & 64.3 & {-} & {-}\\
    \midrule
    AudioSep (16kHz)&5.71&6.60&5.68&5.79 &26.1&29.7&31.5&78.9&68.5&1.21&1.32 \\
    FlowSep (16kHz)      & {-16.62} &  {-16.42}&{-16.65}&{-17.22} & 26.9 &  31.3 &27.6 &80.1&61.4&0.90&1.57\\
    Hybrid-Sep (16kHz) & \textbf{8.82} & \textbf{8.65} & \textbf{8.79} & \textbf{7.84} & \textbf{27.6} &\textbf{32.7} & \textbf{32.8} & \textbf{81.1} & \textbf{79.9} & \textbf{0.85} & \textbf{1.09} \\
    \midrule
    No Processing (44.1kHz) & 0.03 & 0.91&{-}&{-} & 23.2 & 22.7 & 22.3& 71.3  & 64.3 & {-} & {-}\\
    \midrule
    Hybrid-Sep(44.1kHz) &\textbf{8.77} &\textbf{8.62}&\textbf{8.74}&\textbf{7.71}&\textbf{27.3}&\textbf{33.1}&\textbf{33.2}&\textbf{80.4}&\textbf{80.9}&\textbf{0.85}&\textbf{1.09} \\
    Hybrid-Sep w/o SSL (44.1kHz)&8.34&7.07&8.31&6.16&27.1&32.6&28.6&79.3&79.5&1.07&1.29 \\
    Hybrid-Sep w/o CD (44.1kHz)&8.74  &8.53&8.71&7.62&27.2&32.7&29.8&79.9&80.3&1.04&1.25 \\
    Hybrid-Sep 1 stage (44.1kHz)&6.49 &6.29&6.46&5.35&24.6&30.9&27.5&78.6&72.7&1.18&1.29 \\
    
\bottomrule
\end{tabular}
\end{table*}
Signal-to-distortion ratio (SDR) and SDR improvement (SDRi) are employed to evaluate the acoustic quality of separated audio, while CLAPscore, CLAPscoreA, and Fréchet Audio Distance (FAD) measure the semantic accuracy of the separation results. As detailed in \cref{tab:table1}, Hybrid-Sep outperforms all baselines across both acoustic and semantic metrics on all testing datasets.

The proposed ACT method effectively addresses the low-SDR issue observed in diffusion-based LASS models like FlowSep. Compared to the 1-stage model which directly uses text embeddings as queries, the 2-stage design—leveraging audio embeddings as an intermediate representation achieves more than 2 dB SDR improvement. The introduction of consistency loss yields significant gains in semantic evaluation metrics.

Furthermore, incorporating the pre-trained SSL model enhances performance across all datasets, particularly for keyword-based data. The SSL model’s ability to extract high-level semantic features, thereby facilitating acoustic feature reconstruction in the acoustic separation module (ASM). Notably, BEATs is pre-trained on AudioSet which make a superior performance on keyword-audio (KA) test samples derived from the same domain.

\subsection{Case study}
\begin{figure}[t]
  \centering
  \centerline{\includegraphics[width=\columnwidth]{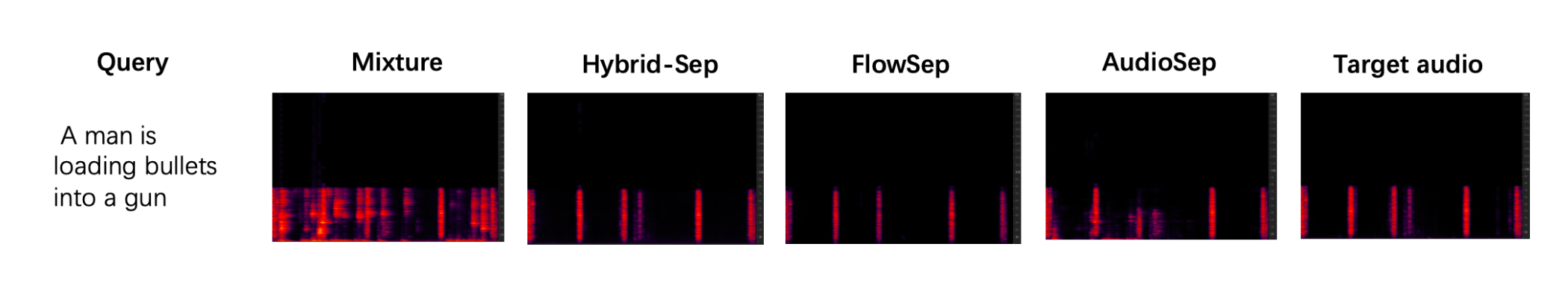}}
  \caption{Case study of different LASS models' performance in DS-S testing dataset}
  \label{fig:case1}
\end{figure}

A case study of different LASS models' performance on the DS-S test dataset is shown in \Cref{fig:case1}. For diffusion-based FlowSep, the separation result is clean but suppresses some parts of the target audio. AudioSep's result has more remaining noise and frequency distortion. Hybrid-Sep provides clean audio separation with less distortion.

\begin{figure}[t]
  \centering
  \centerline{\includegraphics[width=\columnwidth]{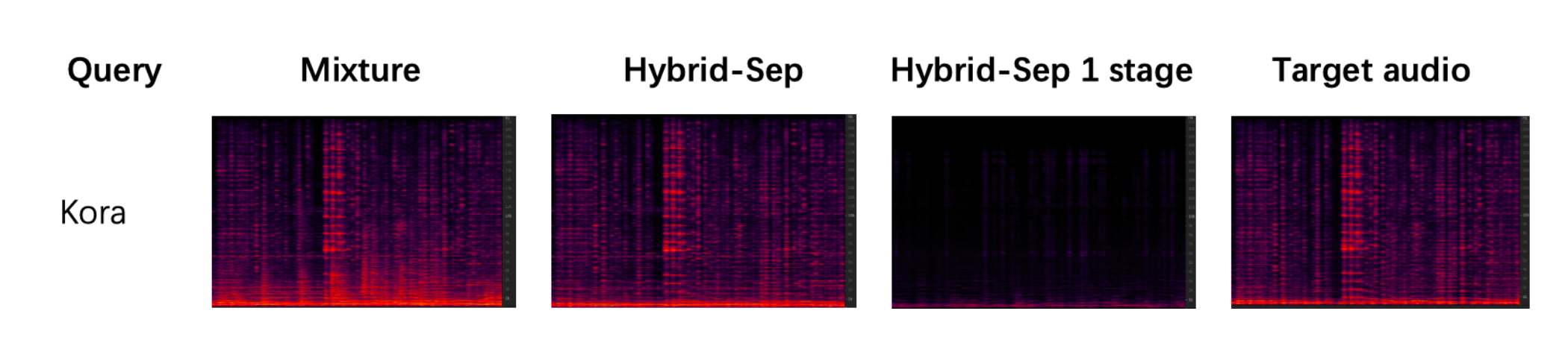}}
  \caption{Case study of an audio source trained with small samples}
  \label{fig:case2}
\end{figure}

The data imbalance issue in audio sources is almost inevitable in LASS model training. We use the example of "Kora" to illustrate few-shot learning performance. During training, only 3 samples containing "Kora" exist in the training dataset. The spectrograms of Hybrid-Sep with 1-stage and 2-stage configurations are shown in \Cref{fig:case2}. As a result, the 2-stage Hybrid-Sep can accurately separate the few-shot trained audio source, while the 1-stage version erroneously suppresses the target audio.


\section{Discussion}
Notice that the CLAPscore measures the similarity between audio and query embeddings, and the value of which is much lower compared with CLAPscoreA. The 1st prize model\cite{lee2024performance} in DCASE LASS Challenge with AudioSep architecture achieved 8.61 dB of SDR in DE-S dataset through caption augmentation. In contrast, Hybrid-Sep attained superior performance (8.82 dB SDR on DE-S) without employing caption augmentation techniques. This suggests that although CLAP has synchronized the language and audio representation, the language-based query still requires additional projection model to ensure the robustness of language-audio semantic consistency. 

As shown in \Cref{fig:case1} and \Cref{fig:case2}, Hybrid-Sep is able to deal with different sampling rate mixtures. Since the sampling rate for SSL and CLAP does not support full-band signals, the performance may still be limited. The sampling rate of T-CLAP encoder is 32 kHz; for signals with only higher frequencies than 16 kHz, the model will lose the separation ability. A full-band CLAP model may be worth investigation to further improve the high-frequency corner cases such as mechanical noise.

BEATs used in Hybrid-Sep is trained with a 16 kHz sampling rate audio and utilize AudioSet only as the training data, which may limit its performance in dealing with open-world 44.1 kHz LASS tasks. It will be a future work to investigate the effect of sampling rate and scaling of datasets in SSL model training.

\section{Conclusion}
This work addresses the critical challenge of aligning linguistic queries with acoustic features in Language-queried Audio Separation (LASS) by proposing HybridSep, a two-stage framework that integrates self-supervised learning (SSL) acoustic representations and Contrastive Language-Audio Pretraining (CLAP) semantic embeddings. Through the novel Adversarial Consistent Training (ACT) strategy, which combines adversarial training with diffusion-based regularization, our method significantly enhances separation fidelity while maintaining semantic alignment.

Experimental results demonstrate that HybridSep outperforms state-of-the-art baselines (e.g., AudioSep, FlowSep) across both acoustic and semantic evaluation metrics. Notably, the two-stage design—utilizing audio embeddings as intermediate queries—achieves both semantic and acoustic improvement compared to direct text-embedding approaches. The inclusion of SSL models further enhances performance, especially for keyword-based tasks, by leveraging high-level semantic features for acoustic reconstruction.

Future work should investigate scaling SSL training to broader sampling rates (e.g., 44.1 kHz) and domains beyond AudioSet, as well as developing full-band CLAP models to address high-frequency edge cases. These extensions could further advance the robustness of language-guided audio separation systems in real-world scenarios.



\clearpage
\bibliographystyle{IEEEtran}
\bibliography{refs25}







\end{document}